# Ferromagnetism Induced by Uniaxial Pressure in the Itinerant Metamagnet $Sr_3Ru_2O_7$

Hiroshi Yaguchi [a], Robin S. Perry [a,b*], and Yoshiteru Maeno [a,b]

[a] Department of Physics, Graduate School of Science, Kyoto University, Kyoto 606-8502, Japan
[b] International Innovation Center, Kyoto University, Kyoto 606-8501, Japan

**Abstract.** We report a uniaxial-pressure study on the magnetisation of single crystals of the bilayer perovskite $Sr_3Ru_2O_7$, a metamagnet close to a ferromagnetic instability. We observed that the application of a uniaxial pressure parallel to the c-axis induces ferromagnetic ordering with a Curie temperature of about 80 K and critical pressures of about 4 kbar or higher. This value for the critical pressure is even higher than the value previously reported (~ 1 kbar), which might be attributed to the difference of the impurity level. Below the critical pressure parallel to the c-axis, the metamagnetic field appears to hardly change. We have also found that uniaxial pressures perpendicular to the c-axis, in contrast, do not induce ferromagnetism, but shift the metamagnetic field to higher fields.

There has been increasing interest in the family of Ruddlesden-Popper type ruthenates $Sr_{n+1}Ru_nO_{3n-1}$ ($n \geq 1$) since the discovery of superconductivity [1] involving spin-triplet pairing in $Sr_2RuO_4$ ($n = 1$). $SrRuO_3$ ($n = \infty$) shows a ferromagnetic metallic ground state with a Curie temperature of 160 K [2]. In this context, $Sr_3Ru_2O_7$ ($n = 2$) is intermediate and an itinerant paramagnet close to a ferromagnetic instability [3]. In fact, metamagnetism has been reported in $Sr_3Ru_2O_7$ [4] and its metamagnetism field is 7.8 T and 5.6 T for $H // c$ and $H // ab$, respectively [5]. Importantly, with improving sample quality [6], the H-T phase diagram in the vicinity of the quantum critical (end) point associated with the metamagnetic transition [7] has been revealed to be more complex and richer [8]. Also ferromagnetism is known to be induced by a uni-axial pressure along the c-axis [9].

We investigated uniaxial pressure effects on the magnetism of $Sr_3Ru_2O_7$. The crystals used were chosen from two batches (C667 and C634), grown by a floating-zone method [6], with different magnetic impurity levels. We used a uni-axial pressure cell with a SQUID (superconducting quantum interference device) magnetometer equipped with an automated background subtraction programme (MPMS, Quantum Design). The cell is piston-cylinder type, and is made of CuBe apart from the cylinder being made of oxygen-free copper to reduce the background signal. Applied pressures were determined from the force applied to the samples at room temperature, which was confirmed to show a reasonable agreement with low-temperature pressure determined by the superconducting transitions of tin and lead [10].

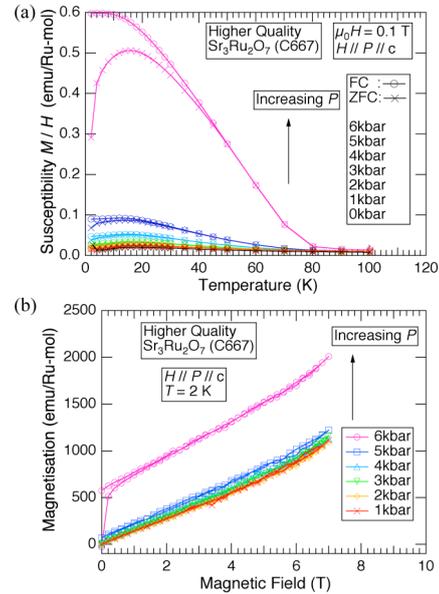

**FIGURE 1.** (a) Temperature dependence of the magnetic susceptibility at 0.1 T and (b) M-H curve at 2 K for the higher quality $Sr_3Ru_2O_7$ under uni-axial pressure along the c-axis.

Figure 1 shows the temperature dependence of the magnetic susceptibility at 0.1 T and the *M-H* curve at 2 K for the higher quality $Sr_3Ru_2O_7$ (C667) with a residual resistivity $\rho_0$ below 1 μΩcm, typically about 0.5 μΩcm, under uni-axial pressure along the c-axis. Clearly, ferromagnetic ordering is induced at a critical pressure between 5 and 6 kbar. The Curie temperature is about 80 K, at which a very abrupt increase in the magnetisation occurs. We have also made measurements on the lower quality $Sr_3Ru_2O_7$ (C634) with $\rho_0$ of 2-3 μΩcm and observed ferromagnetism induced with a considerably lower critical pressure of ~ 4 kbar and a similar Curie temperature of about 80 K, as shown in Fig. 2. In either case, the metamagnetic field is barely affected below the critical pressure, so that the pressure-induced ferromagnetic transition appears to be first order. In a previous report [9], ferromagnetism was induced by a uni-axial pressure along the c-axis with a critical pressure of about 1 kbar, which is much lower than the values obtained in the present study. Taking these facts together, it is inferred that the critical pressure is rather sample-dependent and possibly very sensitive to the impurity level.

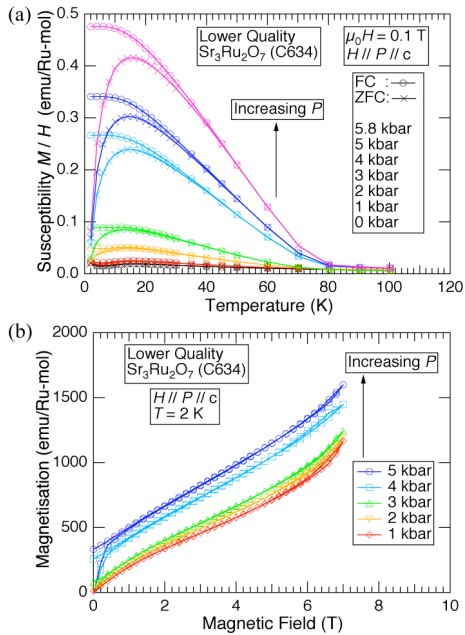

**FIGURE 2.** (a) Temperature dependence of the magnetic susceptibility at 0.1 T and (b) *M-H* curve at 2 K for the lower quality $Sr_3Ru_2O_7$ under uni-axial pressure along the c-axis.

We have also briefly investigated the effects of uniaxial pressure perpendicular to the c-axis using another sample from the higher quality $Sr_3Ru_2O_7$ (C667). The direction of the applied field is different, so that the metamagnetic field at zero pressure in this configuration is consistent with the reported value of about 5.6 T [5]. Figure 3 shows that, in contrast to uniaxial pressure along the c-axis, the metamagnetic field moves substantially to a higher magnetic field.

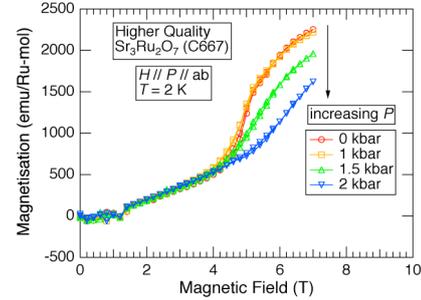

**FIGURE 3.** *M-H* curve at 2 K for the higher quality $Sr_3Ru_2O_7$ under uni-axial pressure parallel to the ab-plane.

In summary, uniaxial pressure parallel to the c-axis induces ferromagnetic ordering with a Curie temperature of about 80 K and critical pressures considerably higher than the value reported (~0.1 kbar) [9], which might be attributed to the difference of the impurity level. Below the critical pressure parallel to the c-axis, the metamagnetic field appears to hardly change. In contrast, uniaxial pressure perpendicular to the c-axis does not induce ferromagnetism, but shifts the metamagnetic field towards higher fields.

## ACKNOWLEDGEMENTS

We thank Dr. N. Takeshita for technical advices and valuable discussions. This work was supported in part by a Grant-in-Aid for Scientific Research from the Japan Society for the Promotion of Science and 21 COE program on "Center for Diversity and Universality in Physics" from the MEXT of Japan.